\input phyzzx.tex
\overfullrule = 0pt
\def\ie{{\it i.e.~}}

\def\oh{{\textstyle {1\over 2}}}

\def\text{\textstyle}

\def\N{{$1\over N$}~}

\def\refmark#1{[#1]}
\def\Q{{${\rm QCD}_2$}~}
\def\2d{{two dimensions}~}
\def\sp{\,\,\,\,}
\def\m{{\mu}}
\def\n{{\nu}}

\magnification=\magstep1
\vsize=21.5 true cm
\hsize=15.4 true cm
\predisplaypenalty=0
\abovedisplayskip=3mm plus6pt minus 4pt
\belowdisplayskip=3mm plus6pt minus 4pt
\abovedisplayshortskip=0mm plus6pt
\belowdisplayshortskip=2mm plus6pt minus 4pt
\normalbaselineskip=12pt

\REF\thooft{G. 't Hooft, {\sl Nucl. Phys. }{\bf
B72}, 461 (1974).  }
\REF\growit{D. Gross and E. Witten, {\sl Phys.
Rev.} {\bf D21}, 446(1980). }
\REF\gromen{D. Gross and P. Mende, }
\REF\spec{J. Polchinski, {\sl Phys. Rev. Lett}
{\bf 68}, 1267 (1992);  M. Green, QMW-91-24
(1991)}
\REF\kazkos{ D. Weingarten, {\sl   Phys. Lett.
} {\bf B90}, 285 (1990); V. Kazakov and I.
Kostov, {\sl   Phys. Lett. } {\bf B128}, 316
(1983); I. Kostov {\sl Phys. Lett.} {\bf B138},
191 (1984); {\sl Nucl. Phys. } {\bf B179}, 283
(1981); K. O'Brien and J. Zuber, {\sl Nucl.
Phys. }{\bf B253}, 621 (1985) }
\REF\kazkostwo{ V. Kazakov and I. Kostov, {\sl
Nucl. Phys.} {\bf B220}, 167 (1983); I. Kostov
{\sl Nucl. Phys. Lett.} {\bf B265}, 223 (1986).
}
\REF\calgro{C. Callan, N. Coote and D. Gross, }
\REF\tobe{ D. Gross,  to be published}
\REF\gropol{D. Gross and A. Polyakov,
unpublished}
\REF\pol{ A. Polyakov, {\sl    Nucl. Phys.}
{\bf B40}, 235 (1982).}
\REF\migdal{A. Migdal, {\sl Zh. Eksp. Teor.
Fiz.} {\bf 69}, 810 (1975)( Sov. Phys. Jetp.
{\bf 42}, 413).}
\REF\rus{B. Rusakov, {\sl Mod. Phys. Lett. }
{\bf A5}, 693 (1990).}
\REF\witetal{ E. Witten, {\sl Comm. Math.
Phys},{\bf 141}, 153(1991; D.Fine, {\sl Comm.
Math. Phys.} {\bf 134}, 273 (1990); M. Blau and
G. Thompson, NIKHEF-H/91-09, MZ-TH/91-17. }
\REF\private{S. Novikov and A. Schwarz, private
communication.}
\REF\tobewat{ D. Gross and W. Taylor, to be
published}
\REF\master{ L. Yaffe, {\sl Rev. Mod.
Phys.}{\bf 54 }, 407 (1982).}
\REF\bary{E. Witten, {\sl Nucl. Phys.} {\bf
B160}, 57 (1979); V. Kazakov and A. Migdal,{\sl
Phys. Lett. } {\bf B103 }, 214 (1981).  }
\REF\matrix{D. Gross and A.Migdal, {\sl Phys.
Rev. Lett.}, {\bf 64}, 717
(1990); M. Douglas and S. Shenker, {\sl Nucl.
Phys.} {\bf B335}, 635 (1990); E. Br\'ezin and
V. Kazakov,
{\sl Phys. Lett.} {\bf 236B}, 144 (1990).}
\REF\induced{ V. Kazakov, and  A. Migdal,
PUPT-1322, May 1992.}
\REF\itz{ Harish-Chandra, {\sl Amer. Jour.
Math.} {\bf 79}, 87 (1957); C. Itzykson and J.
Zuber, {\it Jour. Math. Phys.} {\bf 21}, 411
(1980); J. Duistermaat and G. Heckman, {\sl
Invent. Math.} {\bf 69}, 259 (1982).}
\REF\comments{D. Gross, PUPT 1335, (August
1992).  }
\REF\zn{I. Kogan, G. Semenoff and N. Weiss,
UBCTP 92-022, June 1992.}
\REF\extrinsic{S. Das, A. Dhar, A. Sengupta and
S. Wadia, {\sl Mod. Phys. Lett. } {\bf A5},
1041 (1990); G. Korchemsky, UPRF-92-334
(1992).}
\REF\migmast{ A. Migdal, Princeton preprint
PUPT-1323, June 1992.}
\REF\grokleb{ D. Gross and I. Klebanov,  {\it
Nucl. Phys. \bf B334}, 475 (1990).}
\REF\mignew{A. Migdal, PUPT-1332, LPTENS-92/23
, PUPT-1343 ;
S. Khoklachev and Yu. Makeenko, ITEP-YM-7-92
(August 1992);}

\Pubnum{ LBL 33232\cr  PUPT 1355 }
\date={November, 1992}
\titlepage
\vsize=21.5 true cm
\hsize=15.4 true cm
\title{ {\bf SOME NEW/OLD APPROACHES TO QCD}}
\foot{
Talk delivered at the Meeting on
Integrable Quantum Field Theories, Villa Olmo
and at STRINGS 1992, Rome, September 1992.
This work was supported in part by the Director, Office of
Energy Research, Office of High Energy and Nuclear Physics, Division of
High Energy Physics of the U.S. Department of Energy under Contract
DE-AC03-76SF00098 and in part by the National Science Foundation under
grant PHY90-21984.}
\author{DAVID J. GROSS\
\foot{ On leave from Princeton University, Princeton, New Jersey.} }
\address{LBL}
\vskip 1truein

\abstract{ I discuss some recent attempts to
revive two old ideas regarding an analytic
approach to QCD--the development of a string
representation of the theory and the large $N$
limit of QCD.}

\endpage


\vskip 1in

\centerline{\bf Disclaimer}

\vskip .2in

This document was prepared as an account of work sponsored by the United
States Government.  Neither the United States Government nor any agency
thereof, nor The Regents of the University of California, nor any of their
employees, makes any warranty, express or implied, or assumes any legal
liability or responsibility for the accuracy, completeness, or usefulness
of any information, apparatus, product, or process disclosed, or represents
that its use would not infringe privately owned rights.  Reference herein
to any specific commercial products process, or service by its trade name,
trademark, manufacturer, or otherwise, does not necessarily constitute or
imply its endorsement, recommendation, or favoring by the United States
Government or any agency thereof, or The Regents of the University of
California.  The views and opinions of authors expressed herein do not
necessarily state or reflect those of the United States Government or any
agency thereof of The Regents of the University of California and shall
not be used for advertising or product endorsement purposes.

\vskip 2in

\centerline{\it Lawrence Berkeley Laboratory is an equal opportunity employer.}

\endpage
\chapter{\bf Introduction}

In this  lecture I shall discuss some recent
attempts to revive some old
ideas to address the problem of solving QCD. I
believe that it is timely
to return to this problem which has been
woefully neglected for the last
decade. QCD is a permanent part of the
theoretical landscape and
eventually we will  have to develop analytic
tools for dealing with the
theory in the infra-red. Lattice techniques are
useful but they have not yet
lived up to their promise. Even if one manages
to derive the hadronic
spectrum numerically, to an accuracy of 10\% or
even 1\%, we will not be truly satisfied unless
we have some analytic understanding of the
results.
Also, lattice Monte-Carlo methods can only be
used to answer a small set of
questions. Many issues of great conceptual and
practical interest--in
particular the calculation of scattering
amplitudes, are thus far beyond lattice
control.
Any progress in controlling QCD in an explicit
analytic fashion would
be of great conceptual value. It would also be
of great practical aid to
experimentalists, who must use rather ad-hoc
and primitive models of
QCD scattering amplitudes to estimate the
backgrounds to interesting
new physics.

 I will discuss  an attempt to derive a string
representation of QCD and a revival of the
large $N$ approach to QCD.
Both of these ideas have a long history, many
theorist-years have been
devoted to their pursuit--so far with little
success. I believe that it is
time to try again. In part this is because of
the progress in the last few
years in string theory.  Our increased
understanding of string theory
should make the attempt to discover a stringy
representation of QCD
easier, and the methods explored in matrix
models might be employed
to study the  large N  limit of QCD. For both
political and intellectual
reasons I fervently urge string theorists to
try their hand at these tasks.

\chapter{\bf QCD as a String Theory}
 It is an old idea that QCD might be
represented as a string theory. This
notion dates back even before the development
of QCD. Indeed, string
theory itself was stumbled on in an attempt to
guess  simple
mathematical representations of strong
interaction scattering
amplitudes which embodied some of the features
gleamed from the
experiments of the 1960's. Many of the
properties of hadrons are
understandable if we picture the hadrons as
string-like flux tubes. This
picture is consistent with linear confinement,
with the remarkably
linear Regge trajectories  and with the
approximate duality of hadronic scattering
amplitudes.

Within QCD itself there is internal,
theoretical support for this idea.
First, the \N expansion of weak coupling
perturbation theory can be
interpreted as corresponding to an expansion of
an equivalent string
theory in which the string coupling is given by
\N. This is the famous
result of 't-Hooft's analysis of the \N
expansion of perturbative
QCD \refmark\thooft . The same is true for any
{\em matrix model}--\ie a model invariant under
$SU(N)$ or $U(N)$, in which the basic dynamical
variable is a matrix in
the adjoint representation of the group. The
Feynman graphs in such a
theory can be represented as triangulations of
a two dimensional
surface. This is achieved by writing the gluon
propagator as a double
index line and tiling the graph with plaquettes
that cover the closed
index loops. 't-Hooft's principal result was
that one can use \N to pick
out the topology, \ie the genus=number of
handles, of the surface,
since a diagram which corresponds to a genus
$G$ Riemann surface is
weighted by $({1\over N})^{2G-2}$. The leading
order in  the expansion of the free energy in
powers of
\N is proportional to $N^2$  (reasonable since
there are $N^2$
gluons,  and is given by the {\em planar}
graphs of the theory.

Another bit of evidence comes from the strong
coupling lattice
formulation of the theory. The  strong coupling
expansion
of the free energy can indeed be represented as
a sum over surfaces
\refmark\kazkos .
Again there is  a natural large $N$ expansion
which picks out definite
topologies for these surfaces. This result  is
an existence proof for a
string formulation of QCD. However, the weights
of the surfaces are extremely complicated and
it is not at all clear how to take the
continuum limit.\foot{ There is also the
problem that for large $N$ there
is typically a phase transition between the
strong and weak coupling
regimes \refmark\growit .}

{}From quite general considerations we expect
that the large $N$ limit of
QCD is quite smooth, and should exhibit almost
all of the qualitative
features of theory. Thus an expansion in powers
of ${1\over 3}$ or
$({1\over 3})^2$ might be quite good. The
longstanding hope has been
to find an equivalent (dual) description of QCD
as some kind of string
theory, which would be useful in to calculate
properties of the theory
in the infrared.

The problems with this approach are many.
First, if QCD is describable as a string theory
it is not as simple a theory as that employed
for critical strings. It appears to be easier
to guess the string theory of everything than
to guess the string theory of QCD.
Most likely the
weights of the surfaces that one would have to
sum over will depend on the
{\em extrinsic geometry} of the surface and not
only its intrinsic
geometry. We know very little about such string
theories.
Also there are reasons to believe that a string
formulation would
require many (perhaps an infinite) new degrees
of freedom in
addition to the coordinates of the string.
Finally, there is the important
conceptual problem--how do strings manage to
look like particles at
short distances. The one thing we know for sure
about QCD is that at
large momentum transfer hadronic scattering
amplitudes have
canonical powerlike behavior in the momenta, up
to calculable
logarithmic corrections. String scattering, on
the other hand, is
remarkably soft. Critical string scattering
amplitudes have, for
large momentum transfer, Gaussian fall-off
\refmark\gromen.  How do QCD strings avoid
this?\foot{ Recently there have been some
interesting speculations regarding this problem
\refmark\spec .}\bigskip

 \chapter{\bf   Two Dimensional QCD }

Two dimensional QCD (${\rm QCD}_2$) is the
perfect testing ground
for the idea that gauge theories can be
equivalent
to string theory. First, many features of the
theory
are stringier in two than in four  dimensions.
For example, linear confinement is a
perturbative
feature which is exact at all distances. Most
important
is that the theory is exactly solvable.
This is essentially because in two dimensions
gluons have no physical, propagating
degrees of freedom, there being no transverse
dimensions. In fact \Q is the next best
thing to a topological field theory. The
correlation
functions in this theory will depend, as we
shall see,
only on the {\em topology } of the manifold on
which formulate the theory and on its {\em
area}.
For this reason we will be able to solve the
theory
very easily and explicitly.

Consider for example the expectation value of
the Wilson loop for pure \Q,
$\Tr_R P e^{\oint_C  A_\m dx^{\m}}$,
for any contour,
$C$, which does not intersect itself. Choose
an axial gauge, say $A_1=0$, then the
Lagrangian is quadratic, given by $ \oh \Tr
E^2$,
where $E = \partial_1 A_0$ is the electric
field.
The Wilson loop describes  a pair of charged
particles
propagating in time. This
source produces, in two dimensions, a constant
electric field. The
Wilson loop is then given by the exponential
of the constant energy of the pair
integrated over space and time. This yields,
$$\Tr_R P e^{\oint_C  A_\m dx^{\m}}=
e^{- g^2 C_2(R) A} \sp ,\eqn\aa $$
where $g$ is the gauge coupling, $C_2(R)$
the quadratic Casimir operator for
representation
$R$ and $A$ the area enclosed by the loop. The
expectation value of more complicated
Wilson loops that do self intersect can
also be calculated.  Kazakov and Kostov
worked out a set of rules for such loops
in the large N limit \refmark\kazkostwo. They
are quite complicated.
\Q with quarks is also  soluble, at least in
the
large $N$ limit. The meson spectrum was solved
for $N \to \infty$ by 't Hooft. It consists of
an infinite
set of confined mesons with  masses $m_n$
 that increase as $m_n^2 \sim n$. This provides
one
with a quite realistic and very
instructive  model of quark confinement
\refmark\thooft, \refmark\calgro .

Is \Q describable as a string theory? The
answer is not known, although there is
much evidence that the answer is yes.
I shall describe below a study that I have
carried out
to investigate this issue \refmark\tobe.

 To simplify matters I shall discard the quarks
and consider the pure
gauge theory. This would correspond to a theory
of closed strings only,
quarks are attached to the ends of open
strings. We shall consider the
partition function for a $U(N)$ or $SU(N)$
gauge theory, on an arbitrary
Euclidean manifold ${\cal M}$,
$${\cal Z}_{\cal M} = \int[{\cal D} A^\m]
e^{- {1\over 4 g^2} \int_{\cal M} d^2 x\sqrt{g}
\Tr F^{\m \n}  F_{\m \n}}\sp . \eqn\bb $$
One might  think that in the absence of quarks
the theory is totally
trivial, since in two dimensions there are no
physical gluon degrees of
freedom. This is almost true, however the free
energy of the gluons will
depend non-trivially on the manifold on which
they live. In fact, one
cannot simply gauge the gluons away. If, for
example, ${\cal M}$
contains a non-contractible loop $C$, then if
$\Tr  P
e^{\oint_C  A_\m dx^{\m}}\neq 1 $, one can not
gauge $A_\m$ to zero
along $C$. Thus, the partition function will be
sensitive to the {\em
topology} of  ${\cal M}$.

Although non-trivial the theory is extremely
simple, almost as simple as a topological
theory. It is easy to see that
the partition function will only depend on the
topology and on
the area of the manifold ${\cal M}$. This is
because the theory is invariant under all {\em
area preserving diffeomorphisms}. To
demonstrate this
note that the two-dimensional field strength
can be written as
$ F_{\m \n}= \epsilon_{\m \n} f$, where $
\epsilon_{\m \n}$ is the
anti-symmetric tensor  and $f$ a scalar field.
Thus the action is
$ S= \int \Tr f^2 d\m$, where $d\m = \sqrt{g}
d^2x $ is the volume
form on the manifold.  This action is
independent of the metric, except
insofar as it appears in the volume form.
Therefore the theory is
invariant under area preserving diffeomorphisms
($W_{\infty}$). The
partition function can thus  only depend on the
topology and on
the area of the manifold ${\cal M}$,
$${\cal Z}_{\cal M} = {\cal Z}[G, g^2, A,
N]={\cal Z}[G, g^2 A, N] \sp , \eqn\cc$$
where $G$ is the genus of ${\cal M}$.

Now we can state the conjecture that the
logarithm of this partition
function, the free energy, is identical to the
partition function on some
string theory, with target space ${\cal M}$,
where we would identify
the string coupling with \N and the string
tension with $g^2 N$,
$$\ln\bigl({\cal Z}[G, g^2 A, N]\bigr) = {\cal
Z}_{{\rm Target Space}\,\,
{\cal M}}^{\rm String} [g_{\rm st}={1\over N},
\alpha'=g^2 N  ] \sp .
\eqn\dd $$

 As a candidate for the type of string theory I
am thinking of consider
the Nambu action, wherein
$${\cal Z}^{\rm String}_{\cal M}= \sum_{h = \rm
genus} (g_{\rm
st})^{2h-2} \int {\cal D} x^{\m}(\xi) e^{ \int
d^2\xi \sqrt{g}} \sp ,
\eqn\dd $$
where $g$ is the determinant of the induced
metric
$$g= \det[g_{\alpha \beta}] = \det[ {\partial
x^\m \over
\partial \xi_{\alpha}}
{\partial  x^\n \over  \partial \xi_{\beta}}
G_{\m \n}(x)]  \sp , \eqn\ee $$
and $G_{\m \n}(x)$ is the metric on the
manifold ${\cal M}$.
This string theory, when the target space is
two-dimensional, is indeed
invariant under area preserving diffeomorphisms
of the target space.
To see this note that $\sqrt{g} = |{\partial
x^\m \over \partial
\xi^\alpha}| \sqrt{G}$, which is obviously
unchanged by a map $x^\m
\to x'^\m$ as long as $   |{\partial  x^\m
\over \partial
x^{'\n}}| =1$.\foot{ Actually the Polyakov
action with a two-dimensional target space also
has a $W_{\infty}$ symmetry, although is is
realized in a very nonlinear fashion. One might
speculate that this is related to the well
known $W_{\infty}$ symmetry of the $c=1$ string
theory \refmark\gropol.} Unfortunately the only
way we know to quantize this theory is to
transform it into the Polyakov action, which
upon quantization yields the standard non-
critical string \refmark\pol. This is not what
we want to do here, since the resulting theory
is not even Lorentz invariant. Is there another
quantization of the Nambu string that differs
from the Polyakov quantization in two-
dimensions? The answer is not known.

\section{\bf  Evaluation of the Partition
Function}

The partition function for \Q can easily be
evaluated by means of the following idea,
originally due to Migdal \refmark\migdal. The
trick  is to use a particular lattice
regularization of the theory which is both
exact and additive. For the lattice we take an
arbitrary triangulation of the manifold  and
define the partition function as
$$ {\cal Z}_{\cal M} = \int \prod_L d U_L
\prod_{\rm plaq.} Z_P[U_P]
\sp, \eqn\ff $$
where $U_P= \prod_{L\in {\rm plaq.}} U_L$, and
$Z_P[U_P]$ is some appropriate lattice action.
Any action will do as long as it reduces in the
continuum limit to the usual continuum action.
Instead of the Wilson action, $Z_P(U)= e^{-{1
\over g^2}\Tr (U+U^{\dagger})}$, we shall
choose the {\em heat kernel action},
$$ Z_P= \sum_R d_R  \, \chi_R(U_P) e^{-g^2
C_2(R) A_P } \sp ,\eqn\gg
$$
where the sum runs over representations $R$ of
$ SU(N) $ (or $U(N)$),
$d_R $ is the dimension of $R$, $\chi_R(U_P)$
the character of $U_P$
in this representation, $C_2(R)$ the quadratic
Casimir operator of $R$
and $A_P$ the area of the plaquette.

It is easy to see, using the completeness of
the characters to expand about
$Z_P   \buildrel  {U_L \to 1 +i A_\m dx^\m}
\over\to
\sum_R d_R \chi_R(U_P)= \delta(U_P-1) +\dots
$, that in the continuum limit of this theory
reduces to ordinary Yang-Mills theory. What is
special about the
heat kernel action is that it is additive.
Namely, we can integrate over each link on the
triangulation, say $U_1$, which appears in
precisely two triangles,
using the orthogonality of the characters,
 $\int dV \chi_a(XV) \chi_b(V^{\dagger} Y)
={\delta_{ab}\over
d_a} \chi_a(XY) $, and obtain,
$$\int dU_1 Z_{P_1}(U_2 U_3 U_1)
Z_{P_2}(U_1^{\dagger} U_4
U_5)=  Z_{P_1+P_2}(U_2 U_3 U_4 U_5) \sp .
\eqn\ii $$
This  formula expresses the  unitarity of the
action, since in fact
$Z_P(U)= \langle U| e^{-g^2 A \Delta} |1\rangle
$, where $\Delta$ is the Laplacian on the
group.

We can use this remarkable property of the heat
kernel action to argue that the lattice
representation is exact and  {\em independent
of the triangulation}. This is because we can
use \ii~ in reverse to add as many triangles as
desired, thus
going to the continuum limit. On the other hand
we can use \ii~ to reduce the number of
triangles to the bare minimum necessary to
capture the topology of ${\cal M}_G$. A two-
dimensional manifold of genus  $G$ can be
described by a
$4G$-gon with identified sides: $a_1b_1a_1^{-
1}b_1^{-1} \dots a_Gb_Ga_G^{-1}b_G^{-1}$. The
partition function can be written using this
triangulation as,
$${\cal Z}_{{\cal M}_G} = \sum_Rd_R e^{-{g^2}
C_2(R) A} \int
\prod {\cal D} U_i {\cal D} V_i
\chi_R[U_1V_1U_1^{\dagger}
V_1^{\dagger}\dots U_GV_GU_G^{\dagger}
V_G^{\dagger}] \,\, . \eqn\jj $$
We can now evaluate the partition function
using the orthogonality of the characters and
the relation,
$\int {\cal D}U \chi_a[AUBU^{\dagger}] =
{1\over d_a}  \chi_a[A]
\chi_a[B] $, to obtain \refmark\rus,
\refmark\witetal,
$${\cal Z}_{{\cal M}_G} = \sum_R d_R^{2-2G}
e^{-{\lambda A\over N} C_2(R)  } \sp , \eqn\kk
$$
where $\lambda \equiv g^2N$ is kept fixed.
Thus we have   an explicit expression for the
partition function. It depends, as expected,
only on the genus and the area of the manifold.

\section{ \bf The Large N expansion}

 The formula \kk~ for the partition function is
quite complicated, being written as a sum over
all representations of $SU(N)$. The
representations of $SU(N)$ or $U(N)$ are
labeled by the Young diagrams, with $m$ boxes
of length
$n_1 \geq n_2 \geq n_3 \geq \dots n_m\geq 0$.
Such a representation  has,
$$\eqalign{ C_2(R)&=N \sum_{i=1}^m n_i
+\sum_{i=1}^m(n_i+1-2i) ; \cr
d_R &= {\Delta(h )\over \Delta(h^0)}, \sp
h_i=N+n_i-1,\sp
h_i^0=N-i \cr \Delta(h) & =\prod_{1 \leq
i<j\leq N} (h_i-h_j)   \sp . \cr} \eqn\ll $$
 Thus we have a very explicit sum and one can,
in principle, expand each term in powers of \N
and evaluate the sum.

 What do we expect if the string conjecture is
correct? Consider the expansion in powers of \N
of the free energy,
$$\ln[{\cal Z}_{{\cal M}_G}] =
\sum_{g=0}^{\infty} {1\over N^{2g-2}}
f^G_g(\lambda A) \sp . \eqn\nn $$
If this were given by a  sum over maps of a
two-dimensional surface of genus $g$ onto a
two-dimensional surface of dimension $G$ we
would expect that $f^G_g(\lambda A) \sim
({1\over N})^{2g-2} e^{-\lambda A n}$, where
$n$ is the {\em winding number } of the map,
\ie the topological index that tells us how
many times the map $x(\xi)$  covers ${\cal M
}$. This is the integral
of the Jacobian of the map $\xi \to x$, $\int
d^2 \xi \det[{\partial x^\m\over \partial
\xi^i}]$, which differs from the Nambu area,
$\int d^2 \xi |\det[{\partial x^\m\over
\partial \xi^i}]|$, since the surface can fold
over itself.

Now there is a minimum value that $G$ can take,
given the genus $G$ of the target space and the
winding number $n$. Thus for example there are
no smooth maps of a sphere onto a torus or a
torus onto a genus two surface. Similarly there
are no smooth maps of a genus $g$ surface onto
a genus $g$ surface that wind around it more
than once. To get an idea of the bound
consider holomorphic maps, in which case the
Riemann-Hurwitz theorem state states that $2(g-
1) =  2n(G-1) + B$, where $B$ is the total
branching number.
In the case of  smooth maps there seems to be
the following bound \refmark\private,
$$2(g-1) \geq 2n(G-1) \sp . \eqn\aaa$$

Thus if \Q is described by a string theory we
would expect that
$$ f^G_g(\lambda A) =  \sum_n \cases{ 0  & if
$(g-1) < n(G-1)$ \cr
e^{-n\lambda A} \omega^n_g(A) & otherwise \cr}
\sp .\eqn\mm $$
We can use these inequalities as tests of
whether our conjecture is correct.
To do this we need to expand \kk~in  powers of
\N.

The hardest case is that of the sphere ($G=0$),
since
the sum over representations blow up rapidly
and it is not even evident that there exists a
tamed large $N$ expansion. We can break up the
sum in \kk~ into a sum over representations
with $n$ boxes  in the Young  tableaux since,
for large $N$, $C(R_n) \buildrel N \to \infty
\over \to N\sum_i n_i =Nn$.
Thus,
$$ Z_{G=0} =\sum_R d_R^2 e^{-{\lambda A\over N}
C_2(R) }
\buildrel N \to \infty \over \to   \sum_n
\sum_{R_n} d_{R_n}^2 e^{-n\lambda A}(1 +\dots)
\sp .\eqn\pp $$
To evaluate this we need to evaluate the
following sum,
 $ \sum_{R_n} d_{R_n}^2$. This can be done
using a method of discrete
 orthogonal polynomials \refmark\tobe,
yielding,
 $$\sum_{R_n} d_{R_n}^2 =  {N^2 +n -1 \choose n
} \sp .\eqn\ppp $$
Then it follows that,
$$ Z_{G=0} \to \exp[-N^2\bigr[\ln(1-e^{-\lambda
A}) + {2 \lambda Ae^{-2\lambda
A}\over (1-e^{-\lambda A})^2} + \dots \bigl]
+O(N^0) + \dots ] \sp . \eqn\pqp $$
Here there are no constraints implied by the
inequality \mm~, but the structure of the
expansion is very interesting.

The case of the torus, ($G=1$), is some what
simpler.  One can easily derive that (for
$SU(N)$) \refmark\tobe,
$$\eqalign{Z_{G=1}& =\sum_R  e^{-{\lambda
A\over N} C_2(R)
} \to \exp[ -N^0 \ln \eta(-e^{-\lambda A})  +
\cr & {\lambda A\over N^2}
\sum_{n=1}^\infty e^{-n \lambda
A}[\sum_{ab=n}a^2 b +\sum_{ab+cd=n} ac]
+ \dots]   \sp , \cr}\eqn\qq $$
where $\eta(x)= \prod_{n=1}^\infty (1-x^n)^{-
1}$.
This is totally consistent with the bound
$g\geq 1$.

Most interesting  is the case of $G >1$, where
the inequalities are quite stringent. In this
case one can easily derive \refmark\tobe,
$$Z_{G} \to \sum_n ({1\over N})^{2n(G-1)}e^{-
n\lambda A} \sum_{r={\rm rep\,\, of\,} S_n}
\bigl[{n!\over d_r}\bigr]^{2(G-1)} \sp ,
\eqn\rr $$
where the sum is over representations of the
symmetric group $S_n$ and $d_r$ is
the dimension of the r$^{\rm th}$
representation of $S_n$. Not only is this in
total accord with our expectations, but one can
also show that
$\omega^n_g(A)= $, for $g=1 +n(G-1)$, is
precisely the number of topologically
inequivalent maps on the genus $g$ manifold
onto the genus $G$ manifold with winding number
$n$\refmark\tobewat.

So the large N expansion of \Q looks precisely
like what we would expect from string
considerations. What remains to be understood
are the all the rational numbers that appear as
coefficients of the  powers of $e^{-\lambda A}$
and of \N in terms of the counting of maps of
${\cal M}_g$ onto ${\cal M}_G$. Some of these
are understood, but not all. Then it remains to
construct a string action
that reproduces these counting rules.

\chapter{\bf Induced QCD}

\section{\bf The Large N Limit of QCD}

QCD is hard to solve since it is a theory with
no free, adjustable or small parameters. In
pure QCD (no quarks) the only parameter we can
adjust is the number of colors, $N$. Luckily,
in the large $N$ limit   QCD simplifies
enormously, and this limit remains the best
hope to yield an exact or controllable
treatment of the theory. We know that as $N =
\infty$ only planar graphs survive. More
generally we know that in terms of the
appropriate variables the large $N$ limit of
gauge invariant observables is given, for $N =
\infty$ by the {\em master field}, namely a
solution of an appropriate classical equation
of motion \refmark\master. The large $N$ limit
is in the nature of a semi-classical expansion,
with $N$ playing the role of Planck's constant.
Unlike the running coupling $N$ does not vary
with momentum and we expect the large $N$ limit
to be qualitatively correct for all momenta, to
correctly capture the small distance asymptotic
freedom of the theory as well as exhibit
confinement at large distances. In the $N =
\infty$ we should have an infinite spectrum of
stable mesons and glueballs. Even baryons,
bound states of $N$ quarks, are describable, in
this limit, as solitons of the effective
Lagrangian for the master field \refmark\bary.
Thus the hope has survived that we could find
an exact solution of QCD for $N = \infty$,
which would yield the hadronic spectrum, and
would be the starting point for a systematic
large $N$ expansion which could allow us to
calculate scattering amplitudes.

The standard method of solving a theory in the
large $N$ limit is to find an appropriate
saddlepoint for the partition function. In the
case of QCD this is difficult. Consider the
standard (Wilson) lattice formulation of the
theory,
$$ {\cal Z}_{\rm QCD} = \int \prod_L{\cal D}U_L
e^{-\sum_{\rm plaq.}
{N\over g^2(a)} \Tr \bigl[\prod_L U_L + {\rm
h.c.} \bigr]} \sp .
\eqn\rr
$$
The integrand behaves as the exponential of an
action that is of order $N^2$, thus one might
hope to evaluate it by saddlepoint techniques.
However, the measure is also of order $c^{N^2}$
and therefore one must somehow get rid of $N^2$
degrees of integration before this can be done.
The reason QCD is not yet solved in the large
$N$ limit is that no one knows how to reduce
the theory to $N$ variables per site.

Another theory which is also insoluble in the
large $N$ limit is the non-critical string with
$c >1$. Following the recent success of the
matrix model solutions of string theory
\refmark\matrix, we can construct such strings
if we could deal with the large $N$ limit of a
scalar matrix model in $D$ dimensions, say
$$  {\cal Z}_D^{\rm string}  = \int
\prod_i{\cal D}\phi_i
 e^{-N \sum_i  \Tr  U(\phi_i) +N \sum_{i,
\m=1\dots D} \Tr(\phi_i
\phi_{i+\m})} \sp , \eqn\sst $$
which describes a scalar field on a $D$-
dimensional  lattice. The connection with
string theory is made in the usual way, the
Feynman diagrams of the perturbative expansion
of \sst~ correspond, in an expansion in powers
of \N, to triangulations of two-dimensional
surfaces. The scalar fields correspond to
matter on this surface and thus, \sst~, could
yield, at the appropriate critical point where
the mean number of triangles diverges, a $c=D$
string theory. The standard approach to the
large $N$-limit of such a theory is to
diagonalize the matrices $\phi$, \ie to pass to
{\em radial} coordinates,
$ \phi_i = \Omega_i
\lambda_i\Omega^{\dagger}_i$, where $\lambda_i$
is diagonal. In terms of these variables,
$${\cal Z}_D^{\rm string} =\int \prod_i {\cal
D} \lambda_i{\cal D} \Omega_i
\Delta^2(\lambda_i)
e^{-N \sum_i  \Tr  U(\lambda_i)
+N\sum_{<ij>}\Tr\bigl(\lambda_iV_{ij}\lambda_j
V^{\dagger}_{ij}\bigr)}
\sp ,  \eqn\ss
$$
where $V_{ij}= \Omega_i\Omega_j^{\dagger}$ and
$\Delta(\phi)=\prod_{i<j}(\phi_i-\phi_j)$.

The next step is to integrate out the
diagonalization matrices, $\Omega_i$.
We can change variables from the
$\Omega_i's$,defined on the sites to the
$V_{ij}$'s, defined on the links
$\prod_i {\cal D} \Omega_i =\prod_{<ij>}{\cal
D}V_{ij}\prod_{\rm
plaq.} \delta(1-\prod_L V_{ij}) .$ The
constraints arise since the  $V_{ij}$'s are
pure gauge fields. If not for the constraints
we could perform the integral over the
$V_{ij}$'s and reduce the integral to one over
$N$ variables per site that could be evaluated
by saddlepoint techniques. It is these
constraints that have prevented the
construction of strings with $c>1$.

Now let us combine these two models to consider
QCD with adjoint scalar matter,
$$ \eqalign{{\cal Z}_{\rm QCD}^{\rm adj}& =
\int \prod_L{\cal D}U_L
\prod_i{\cal D} \phi_i
e^{-N \sum_i  \Tr  U(\phi_i) +N \sum_{i,
\m=1\dots D}
\Tr(\phi_i U_\m \phi_{i+\m}U^{\dagger}_\m)} \cr
& e^{-{N\over g^2(a)}\sum_{\rm plaq.}\bigl[ \Tr
(\prod_L U_L + {\rm
h.c.})\bigr]} \sp .  \cr }   \eqn\tt  $$

This theory is invariant under standard gauge
transformations,
$\phi_i\to V_i \phi_iV_i^{\dagger} ; \sp U_\m
\to V_i U_\m
V^{\dagger}_{i+\m} , $which allow us to
diagonalize the $\phi$'s. However
the presence of the Wilson action prevents us
from handling this theory for large $N$.
If set the gauge coupling to zero, we recover
the previous model, since in this limit we can
drop the Wilson action term, as long as we
enforce the constraints, $\tr[U_P]=1$. However
if we take the opposite limit, \ie set
$g=\infty$, then we can simply drop the Wilson
action  and the model will be soluble in the
large $N$ limit. This is {\em induced QCD}
\refmark\induced.

Induced QCD has the one great advantage of
being soluble, or at least reducible to a well
defined master field equation. This is because
the integral over the link matrices can now be
performed. This is the famous {\em Itzykson-
Zuber} integral \refmark\itz,
$$I(\phi, \chi) \equiv \int {\cal D} U e^{N
\Tr\bigl[ \phi U \chi
U^{\dagger}\bigr] }= {\det \bigl[
e^{N\phi_i\chi_j}\bigr]\over
\Delta(\phi)\Delta(\chi) } \sp . \eqn\uu
$$
This formula is very profound, underlies all
the analysis of the $c=1$, matrix model, and
can be derived in many ways. One is the
demonstration that the integral is given
exactly by the WKB approximation, and the
answer is simply the sum over the $N!$
saddlepoints, for which are the $U$ are
permutation matrices.

Although soluble this model appears to be very
far from QCD, since asymptotic freedom
instructs us to set the lattice coupling to
zero, not  infinity, in the continuum limit.
However, Kazakov and Migdal argued that even
though there is no kinetic term for the gauge
field, it could be {\em induced} at large
distances \refmark\induced. They argued that if
one integrates out the scalar mesons (even in
the case of noninteracting scalars with
$U(\Phi) = \oh m^2 \Phi^2$),
then at distances large compared to $a$, one
would induce in four
dimensions an effective gauge interaction,
$$ S_{\rm eff} (U) \sim {N\over 96 \pi^2}
\ln({1\over m^2 a^2}) \Tr
F_{\m\n}^2 + {\rm finite\,\, as\,\, }a\to 0
\sp . \eqn\bbb $$
This is simply the one loop vacuum graph
for the scalars in a background gauge field,
which is logarithmically divergent in four
dimension. Now this looks very much like the
ordinary Yang-Mills action, ${1\over g^2(a)}\tr
F_{\mu \nu} F^{\mu \nu}$, if we recall that
asymptotic freedom tell us that  ${1\over
g^2(a)} = {11N\over 48 \pi^2} \ln({1\over
M_g^2a^2}),$ where $M_g$ is a mass scale for
QCD, say the glueball mass. We can therefore
identify these two (the fact that
there are $N^2$ scalars is crucial, as is the
sign of the effective action which is due to
the non-asymptotic freedom of the scalars.)
If we do so then we find that, $M_g^2 =
m^{23\over 11} a^{1\over 22} . $
Thus in the continuum limit the adjoint scalars
become infinitely massive and decouple, but not
before they have drive ${1\over g^2}$ up, from
zero at distance $a$ to the large QCD value
atdistance ${1\over m}$, where  ${1\over M_g}
>> {1\over m}>>a .$
The basic idea is that the infrared slavery of
the scalars, at the
size of the lattice spacing, produces an
effective
gauge theory at a larger scale (much larger
than the inverse scalar mass), which then
produces
the usual asymptotically free fixed point
theory.

There are many problems with this idea. For one
the hard gluons are
not absent and their contribution will
overwhelm  that of the
scalars at short distances. Their
asymptotic freedom is  more powerful than the
infrared slavery of
scalars.
Another issue is that the above theory
possesses a much larger symmetry than the
$SU(N)$ gauge symmetry of the usual lattice
action. It is not difficult to see that, in D
dimensions, it is invariant under $(D-1)\times
(N-1)$ extra local $U(1)$-gauge    symmetries.
This is because the transformation
$U_{\mu}(x)\!\! \to
V^{\dagger}_{\mu}(x)U_{\mu}(x)V_{\mu}(x+\mu
a)$,
leaves the action invariant as long as
$V_{\mu}(x)$ is   a unitary matrix
that commutes with $\Phi(x)$.  If  $V_{\mu}(x)$
were independent of
$\mu$ then this would be the ordinary gauge
invariance. Thus we have $D-1$
new gauge symmetries, which are of course
isomorphic to the special unitary
transformations that commute with $\Phi$
\refmark\comments. Thus  $V_{\mu}(x)=
D_{\mu}(x)\Omega (x)$, where $\Omega (x)$ is
the unitary matrix that
diagonalizes $\Phi$ and $D_{\mu}(x)$ is
diagonal.

A subset of this symmetry is the, field
independent, local $Z_N$ symmetry, $U_{\mu}(x)
\!\! \to Z_\mu  U_{\mu}(x)
Z_\mu^{\dagger}$, where $Z_\mu$ is an element
of the center  of the group.
This symmetry alone prevents the Wilson loop
from acquiring an expectation value. A Wilson
loop contains different links, and thus
$W(C) = \langle \prod_{L\in C}U_L\rangle \to
\bigl(\prod_L
Z_\m\bigr)  W(C) \Rightarrow W(C)=0.$ This
symmetry must be broken if we are to recover
the QCD fixed point from this formulation
\refmark\zn.

Finally, as we shall see, the simple Gaussian
model is soluble and the answer is very simple
and {\em not} equivalent to QCD
\refmark\comments. However, there are
interesting attempts to save the model and
furthermore even if it does not yield a
solution of QCD it might provide some
interesting soluble matrix models which could
yield new solutions of new string theories.
Induced QCD is a  matrix model and thus it
corresponds to some kind of sum over surfaces.
If we look at the
Itzykson-Zuber integral we note that it could
be expressed as,
$$I(\phi,\chi) =\exp\bigl[\oh \Tr\phi^2
\Tr\chi^2 + a \Tr\phi^4
\Tr\chi^4 + {b\over N^2} (\Tr\phi^2)^2(
\Tr\chi^2)^2 + \dots\bigr]
\sp . \eqn\qwq $$
These terms will affect the structure of the
large $N$ expansion of the Feynman diagrams,
and can be interpreted as yielding extra
weights when the two-dimensional surfaces
intersect \refmark\extrinsic. Thus this model
corresponds, perhaps, to some kind of string
theory with weights that depend on the
extrinsic geometry.

\section{\bf Solution of the Gaussian model}

To try to solve the model of induced QCD we
first integrate out he $U_L$'s, then look for
extrema of the effective action,
$$S[\phi_i] = N^2\bigl[ {1\over N} \Tr \sum_i
U(\phi_i) +{1\over N^2}\sum_{i,\m}\ln
I(\phi_i,\phi_{i+\m}) + {1\over N^2}  \sum_i
\ln \Delta^2(\phi_i)\bigr] \sp . \eqn\ab $$
In the large N limit the integral will be
dominated by a
translationally invariant saddlepoint
for the density of eigenvalues  of the matrices
$\Phi_i$,
$\rho(x)\equiv  {1\over N} \sum_{a=1}^N
\delta(x-\phi_a)$. Migdal has derived the
master  field equation for the saddlepoint,
using the Schwinger-Dyson equations that are
satisfied by $I(\phi,
\chi)$ \refmark\migmast. These are consequences
of the fact that $I$ satisfies
$\tr[({1\over N}{\partial \over
\partial\phi})^k] I = \tr (\chi)^k I$.
The net result is that one
derives an   equation  for the function
$F(z)\equiv  \int dz {\rho(\nu) \over z-\nu}$,
whose imaginary part is ${\rm Im} F(\nu) = -
\pi \rho(\nu)$,
$$
{\rm Re}F(\lambda)  =
P \int {d\nu \over 2 \pi i} \ln [{ { \lambda -
 {1 \over 2D} U'(\nu) - {D-1 \over D} {\rm Re}
F(\nu)+i\pi \rho (\nu)} \over
 { \lambda -   {1 \over 2D} U'(\nu)- {D-1 \over
D} {\rm Re} F(\nu) -i\pi \rho (\nu)} }  ].
\eqn\ggg $$
This equation is much more complicated than the
usual Riemann-Hilbert problem that one obtains
for simple matrix models. It is sufficiently
non-linear and complex that one might imagine
that it describes QCD.

The master field equation simplifies
dramatically for $D=1$. This is because in one
dimension the gauge field can be gauged away
completely, thus the model is equivalent to a
scalar field on a one-dimensional lattice. The
large  $N$ limit of this model describes the
$c=1$ string on a discrete target space, a
model which has been solved in the double-
scaling limit for small lattice spacing
\refmark\grokleb. It undergoes a phase
transition at a finite lattice spacing and it
might be very instructive to use \ggg~ to
explore this phenomenon.

In
particular for the quadratic potential the path
integral is  Gaussian,
$$Z= \int \prod_n {\cal D} \Phi_ne^{ -  N
\sum_n  {\rm Tr} \{  {m^2\over 2} \Phi_n^2-
\Phi_n \Phi_{n+1} \}}\,\,.  \eqn\mm
$$
Thus the eigenvalues of $\Phi$ will be given by
the semi-circular distribution,
namely $\pi  \rho(\nu) =  \sqrt{\mu - {\mu^2
\nu^2 \over 4}}$,
where $\mu$ is determined by
the mean of the squares of the eigenvalues,
$\langle {1\over N} {\rm Tr}(  \Phi^2  )\rangle
=
{1\over \mu}$.  It is therefore sufficient to
calculate the expectation
value of  $ {1\over N} {\rm Tr}(   \Phi^2  ) $,
which is given by the one loop integral,
$$ {1\over N} {\rm Tr}(   \Phi^2  )=
\int_{-\pi}^{\pi}  {d p\over 2 \pi} {1 \over
m^2 +2{\rm cos }p  }
= {1\over \sqrt{m^4-4}}\equiv {1 \over\mu}\,\,.
\eqn\nn $$
It is easy to verify that this solves \ggg,
using the fact that
$$ F(z) = { \mu z \over 2} - \sqrt{  {\mu^2 z^2
\over 4}-\mu};
\,\,\, {\rm Re} F(\nu) = \oh \mu \nu , \,
\eqn\jk$$
However, if we return to \gg, we see that the
integral
involved is of the same form for any $D$,
as long as $\Re V'(\nu)$ is linear in $\nu$.
This suggests that we can find a solution of
\ggg~with a semi-circular distribution of
eigenvalues for a quadratic potential in any
dimension \refmark\comments.

Indeed, one can see that a semi-circular
distribution of eigenvalues satisfies \ggg~ for
any $D$ as long as,
$$
\mu_{\pm}(D) = {m^2(D-1) \pm D\sqrt{ m^4 -4(2D-
1)} \over 2D-1 }\,\,\, . \eqn\ppp
$$
This solution is much too trivial to describe
QCD. In particular, for $D>1$ there is no
sensible continuum limit of the model.

Is the solution unique? To see that it is note
that in the master field equation the dimension
of space-time enters only via the number of
nearest neighbors of a given site, the
coordination number of the lattice.\foot{ I
thank C. Bachas for emphasizing this point to
me.} The translationally invariant background
scalar field is the same for any lattice with
the same coordination number. The observables,
say the scalar propagator, will of course
depend on the full structure of the lattice,
but not he background field. Therefore we can
choose another simpler lattice with the same
coordination number, say a Bethe lattice, which
contains no closed loops. For such a lattice,
as in the case of the $D=1$ model, the gauge
field can be eliminated completely, and the
model is equivalent to,
$$ Z_{\rm Bethe  Lattice}= \int {\cal D} \phi_i
e^{-N\sum_i\Tr {m^2\over 2} \phi_i^2 + N
\sum_{<ij>} \Tr[\phi_i \phi_j] } \sp .
\eqn\bethe
$$
This model is easily soluble. We define
$Z(\phi)$ to be the partition function of a
branch of the Bethe lattice with coordination
number $2D$, so that $Z= \int {\cal D} \phi
Z(\phi)^{2D}
e^{-{m^2\over 2} N\Tr \phi^2}$. $Z(\phi)$
satisfies the equation,
$$ Z(\phi) = \int {\cal D} \phi' Z(\phi')^{2D-
1} e^{-{m^2\over 2} N\Tr \phi'^2
+N\Tr[\phi\phi']} \sp . \eqn\beteq $$
These equations are easily soluble. Take
$Z(\phi)$  to have the form $Z(\phi) = ce^{-
N{\alpha \over 2} \Tr \phi^2}$, then \beteq~
determines $\alpha $ to  equal
$ \alpha= {-m^2 \pm \sqrt{m^4-4(2D-1)} \over
2(2D-1)}$. Then \bethe~ can be used to
determine ${1\over N} \Tr \phi^2 = {1 \over
m^2  +2 \alpha D}$, which agrees  precisely
with ${1\over \mu}$ as given by \ppp.

\section{\bf  Prospects}

The simplest Gaussian model fails, but all hope
is not lost. It is certainly possible to induce
QCD if one introduces enough flavors of matter.
The problem is that one then loses solubility.
It might be that the self interactions of the
scalars could be adjusted to drive the theory
towards the asymptotically free fixed point.
This hope has been pursued with great vigor by
Migdal, who has also considered adding
fermions, not too many so that the model
remains soluble, so as to break the $Z_N$
symmetry \refmark\mignew. Time will tell
whether this will succeed.
Even if it does not   these model might yield a
new class of interesting soluble matrix models
which could teach us something about new
classes of strings, perhaps strings that depend
on extrinsic geometry.
For this reason alone it is worth studying
these models.
\refout
\end